\documentclass[sigconf]{acmart}
\AtBeginDocument{%
  }

\setcopyright{acmlicensed}
\copyrightyear{2025}
\acmYear{2025}
\acmDOI{XXXXXXX.XXXXXXX}
\acmConference[IVA '25]{Intelligent Virtual Agents}{September 17--19,  2025}{Berlin, GE}
\acmISBN{978-1-4503-XXXX-X/2018/06}

\usepackage{subfigure}

\begin{document}

\title{Are You Listening to Me? Fine-Tuning Chatbots for Empathetic Dialogue}

\author{Paulo Ricardo Knob}
\email{paulo.knob@edu.pucrs.br}
\orcid{0009-0001-9872-9355}
\affiliation{%
  \institution{Pontifícia Universidade Católica do Rio Grande do Sul}
  \city{Porto Alegre}
  \state{RS}
  \country{Brazil}
}

\author{Leonardo Scholler}
\email{leonardo.scholler@edu.pucrs.br}
\affiliation{%
  \institution{Pontifícia Universidade Católica do Rio Grande do Sul}
  \city{Porto Alegre}
  \state{RS}
  \country{Brazil}
}

\author{Juliano Rigatti}
\email{juliano@comunicacaocomconsciencia.com.br}
\affiliation{%
  \institution{Nelogica Sistemas de Software Ltda}
  \city{Porto Alegre}
  \state{RS}
  \country{Brazil}
}

\author{Soraia Raupp Musse}
\email{soraia.musse@pucrs.br}
\affiliation{%
  \institution{Pontifícia Universidade Católica do Rio Grande do Sul}
  \city{Porto Alegre}
  \state{RS}
  \country{Brazil}
}


\renewcommand{\shortauthors}{Blind}

\begin{abstract}
  Conversational agents have made significant progress since ELIZA, expanding their role across various domains, including healthcare, education, and customer service. As these agents become increasingly integrated into daily human interactions, the need for emotional intelligence, particularly empathetic listening, becomes increasingly essential. 
 In this study, we explore how Large Language Models (LLMs) respond when tasked with generating emotionally rich interactions. Starting from a small dataset manually crafted by an expert to reflect empathic behavior, we extended the conversations using two LLMs—ChatGPT and Gemini. We analyzed the emotional progression of the dialogues using both sentiment analysis (via VADER) and expert assessments. 
 While the generated conversations often mirrored the intended emotional structure, human evaluation revealed important differences in the perceived empathy and coherence of the responses. These findings suggest that emotion modeling in dialogues requires not only structural alignment in the expressed emotions but also qualitative depth, highlighting the importance of combining automated and human-centered methods in the development of emotionally competent agents.
\end{abstract}


\begin{teaserfigure}
  \includegraphics[width=\textwidth]{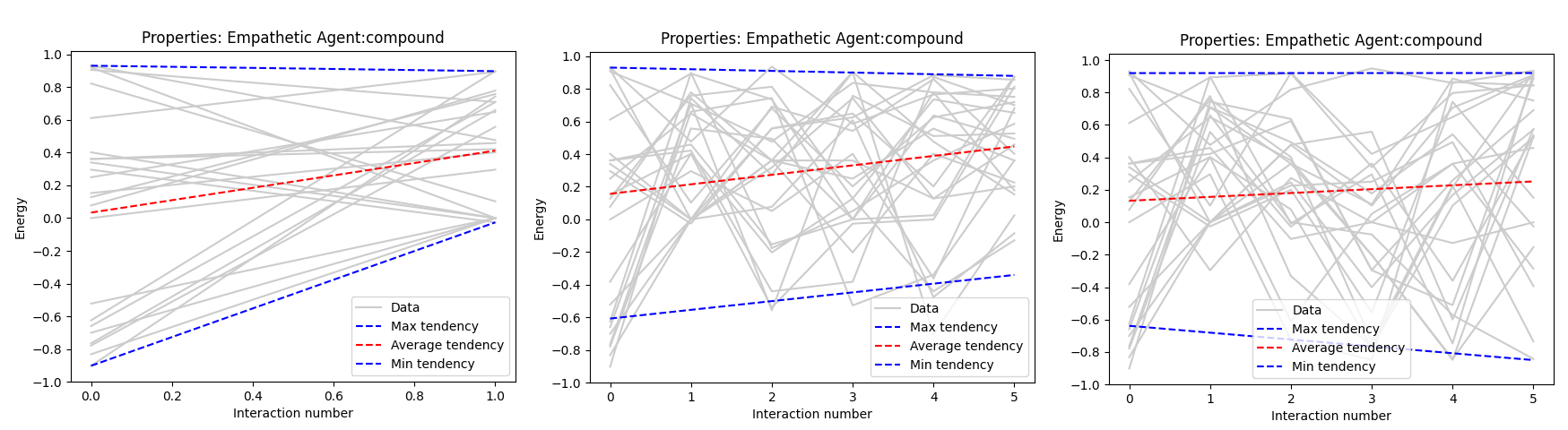}
  \caption{
  Emotional energy trends across three datasets: expert-authored (left), ChatGPT-extended (center), and Gemini-extended (right). The red line shows the average emotional tendency evolution, and the blue lines represent minimum and maximum values. While all three datasets display a general increase in emotional energy, only the expert-authored set reflects a truly empathetic progression. The LLM-generated datasets approximate the pattern, but human evaluation reveals that genuine empathy is not always achieved.}
  \Description{Energy tendency of the compounded interactions for the expert's dataset (left), dataset extended by ChatGPT (middle), and dataset extended by Gemini (right). Y axis shows the energy value (as computed by VADER tool), and X axis shows the flow of the interactions (from the beginning to the end). Blue lines show the maximum and minimum energy tendencies, while the red line represents the average energy tendency. It is possible to notice a growing pattern of the energy tendency, from lower to higher values, as the conversation flowed, which is particularly evident in the red line (Average tendency) of the figures.}
  \label{fig:teaser}
\end{teaserfigure}


\maketitle

\section{Introduction}
As technology evolves, it becomes more integrated with the routine of people. Many tasks that, in the past, required human intervention, can be performed nowadays by specialized computer systems, thanks to the Artificial Intelligence (AI)~\cite{manickam2022artificial}. Since the pioneer work of Joseph Weizenbaum, who proposed the chatbot ELIZA~\cite{weizenbaum1966eliza}, conversational agents have been impacting on how humans interact with machines~\cite{talyshinskii2024potential}. In fact, the advances in technology and, more specifically, in the AI field, led to the development of Natural Generative Language (NGL) models, which include the Generative Pre-trained Transformer (GPT) series~\cite{brown2020language}, released by OpenAI as ChatGPT, and Gemini~\cite{team2023gemini}, released by Google, among others. Such systems are based on Large Language Models (LLMs) and can both emulate the conversation between human beings and generate new content, based on the data in which they were trained~\cite{abd2023large}.

Although LLMs, such as ChatGPT and Gemini, are able to generate coherent information and human-like responses, their ability to identify and describe emotions is uncertain yet~\cite{elyoseph2023chatgpt}. Emotion is closely related with empathy, specially when we think about its affective attributes, which involve physiological enthusiasm and spontaneous affective responses to someone else's display of emotions~\cite{prendinger2005using}. Therefore, the modeling of an empathetic behavior in large language models is a research problem which is open for discussion and ideas.

The goal of this work is to assess whether or not a LLM-powered chatbot can be fine-tuned for empathetic listening, supporting the speaker in understanding feelings and needs for different situations. In order to do so, an experiment was conducted to detect the energy/polarity of the conversation, from the beginning until the end, using two well-known LLM-powered chatbots: ChatGPT, from OpenAI, and Gemini, from Google. Moreover, an expert was asked to evaluate the interactions generated by both chatbots. 
Figure~\ref{fig:teaser} illustrates the evolution of emotional energy throughout the interactions in three different datasets: the original one created by a human expert (left), and the extended versions generated by ChatGPT (center) and Gemini (right). In all three plots, the y-axis represents the energy values (as computed by VADER, ranging from -1 to 1), and the x-axis represents the progression of the conversation. The red curve shows the average emotional trend, while the blue curves indicate the minimum and maximum values at each interaction step. The manually authored dataset exhibits a clear and gradual increase in emotional energy, aligning with the expected behavior of an empathetic listener, helping the speaker move from negative to more positive affective states. The ChatGPT-extended dataset mirrors this upward trend, suggesting that the model can emulate the intended emotional trajectory; however, human evaluation later revealed that these responses often lacked genuine empathy. The Gemini-generated dataset also presents an overall increase in average energy, but with lower minimum values at the beginning and at the end of the interaction, indicating greater difficulty in processing or responding to the speaker’s initial negative emotions. These results highlight that while lexical sentiment analysis captures general emotional patterns, it is not sufficient to assess whether true empathetic listening occurred, underscoring the importance of human evaluation in this context.



\section{Related Work}
\label{sec:relatedwork}

Arjmand et al.~\cite{arjmand2024empathic} introduce the concept of 'empathic grounding' in conversational agents, which includes the listener empathy for the speaker’s affective state. The authors argue that models of empathic grounding for embodied agents should be multi-modal to facilitate natural and efficient communication. Thus, they describe a multi-modal model that takes as input user speech and facial expression to generate multi-modal grounding moves for a listening agent, using a large language model (namely, GPT-3.5). The authors compare their work with one that only generates non-affective grounding cues, and the results indicate that empathic grounding increases user perceptions of empathy, understanding, emotional intelligence, and trust.

Hasan et al.~\cite{hasan2023sapien} present the Synthetic Anthropomorphic Personal Interaction ENgine (SAPIEN), a platform powered by Large-Language Models (LLM), with high-fidelity virtual agents that can engage in real-time open-domain conversations, while also expressing emotions through voice and facial expressions. Their platform allows customization of the virtual agent, tweaking its personality, background, and conversation premise. Besides that, the user can choose to get the conversation analyzed, after the virtual interaction, and receive feedback on their communication skills. The authors also comment about the range of domains in which SAPIEN can be used, such as mental health, communication training, language learning, education, and healthcare. The application is available online.

Elyoseph et al.~\cite{elyoseph2023chatgpt} uses the Levels of Emotional Awareness (EA) Scale to analyze ChatGPT’s responses to twenty scenarios, comparing its EA performance with the general population norms, as reported by a previous study. Additionally, a second examination was performed one month later to measure the improvement over time. In the first examination, ChatGPT demonstrated significantly higher performance than the general population in all scores. In the second examination, ChatGPT’s performance significantly improved, almost reaching the maximum possible scores, with extremely high accuracy levels (9.7/10). Thus, the authors comment that ChatGPT can generate appropriate EA responses, and its performance may improve significantly over time. However, further research is needed in order to better understand the potential benefits and risks, while also refining it to promote mental health.

Bill et al.~\cite{bill2023fine} aim to fine-tune a LMM to act in the psychology area. To do so, the authors use a method known as Reinforcement Learning from Human Feedback (RLHF), which is an approach for reinforcement learning that aims to improve the performance of a given LLM by using the feedback of interacting humans. According to the authors, using such human feedback, the model is able to learn from specific examples and adapt to a particular task. The results achieved presented no significant difference between the pre-trained model and the fine-tuned model, which suggests that the fine-tuned model had little or no difference when compared with the model with no fine tuning.

Chen et al.~\cite{chen2023soulchat} argue that large language models often tend to provide universal advice when applied in the psychological counseling field. However, in such scenarios, the person is usually looking for more elaborated support, such as empathy, trust, understanding, and comfort. With that in mind, the authors built a dataset comprised of over 2 million multi-turn empathetic conversations, focusing on responses that exhibit questioning, comfort, recognition, listening, trust, and emotional support. The dataset was used to fine-tune LLMs, showing that their empathetic response capabilities were significantly enhanced.

\section{Proposed Experiment}
\label{sec:metodologia}

This work aims to assess whether or not a LLM-powered chatbot can be fine-tuned for empathetic listening, supporting the speaker in understanding feelings and needs for different situations. The next sections present more details about the experiment conducted for this end. 
Section~\ref{sec:datasets} shows the configuration of an experiment conducted to detect the energy/polarity of the conversation between two individuals, making use of two different well-known LLM-powered chatbots, namely ChatGPT and Gemini. Section~\ref{sec:expertQuest} presents the configuration of the questionnaire responded to by the expert, based on the interactions generated by both chatbots.

\subsection{Datasets Configuration}
\label{sec:datasets}

The goal of this experiment is to evaluate an LLM chatbot that can conduct a conversation with a person to facilitate empathic listening, thereby creating a sense of comfort in the person. For this experiment, we chose to work with two different chatbots: OpenAI LLM, known as ChatGPT, and Google LLM, known as Gemini. The idea is to utilize techniques of prompt engineering to fine-tune the model and guide it to engage in empathetic listening. For this end, the following steps were conducted:

\begin{itemize}
    \item (1) Gather examples of conversations curated by an expert, to serve as a base and ground truth;
    \item (2) Define a benchmark technique to quantify and classify the responses produced by the model;
    \item (3) Generate the interactions, simulating a real conversation.
\end{itemize}

For the first step, an expert in communication was assigned to manually create 24 texts with example interactions between two people, where the listener acted as an empathetic listener. A dataset was created with those texts. 
The conversations present in this dataset portray interactions between a speaker and a listener, where the listener acts as an empathetic listener. Each of the 24 conversations is comprised of 5 interactions, where the speaker (who starts the conversation) has three interactions, and the listener has 2. Table~\ref{tab:expertDatasetExample} shows an example of a conversation manually tailored by the expert.

\begin{table}[htb]
\centering
\footnotesize
\begin{tabular}{|p{0.1\linewidth}|p{0.8\linewidth}|}
\hline
\textbf{Agent} & \textbf{Utterance} \\ \hline
Speaker & Yesterday I came home and the sink was once again full of dirty dishes. The guy doesn't care. I've already asked him to wash it as he cooks so it doesn't accumulate, but it's no use! I can not take it anymore. \\ \hline

Listener & I'm listening to you and realizing that there's anger, frustration there, does that make sense? \\ \hline

Speaker & Yes, and tiredness as well. \\ \hline

Listener & And perhaps these feelings are present because you need reciprocity and respect. \\ \hline

Speaker & That's it, I really need it. Of consideration as well. \\ \hline
\end{tabular}
\caption{Example of conversation manually tailored by the expert.}
\label{tab:expertDatasetExample}
\end{table}

Then, for the second step, we consider the concept of conversation energy, where negative and positive feelings are evaluated as negative and positive energy, respectively, ranging from -1 (negative) to 1 (positive). To perform this energy evaluation, we utilized the Valence Aware Dictionary and sEntiment Reasoner (VADER) tool~\cite{hutto2014vader}, a lexical and heuristic set that generates a numerical component indicating sentiment, ranging from negative to positive, to describe the energy of the text.

Finally, for the third step, the fine-tuned LLM was used to generate several conversations simulating interactions between two people. In this step, four additional datasets were created: the ChatGPT extended dataset, which was created extending the interactions of the dataset manually tailored by the expert, using ChatGPT; the Gemini extended dataset, which was created in the same way, but using Gemini instead; the ChatGPT control dataset, where ChatGPT generated the conversations entirely, with no fine-tuning concerning empathetic listening; and the Gemini control dataset, which was generated in the same manner, but using Gemini instead. 

The extended datasets (for both ChatGPT and Gemini) were generated by continuing the conversations of the first dataset (manually tailored by the expert). The idea of extending the data to more points also serves to validate the LLM's coherence and cohesion in maintaining longer conversations until a conclusion. The first points were not changed, so the data analysis is preserved in the first two points. For normalization purposes, all conversations were truncated at 12 interactions, where the speaker (who starts the conversation) counts with 6 interactions, and the listener (role played by the LLM chatbot) counts with another 6. Table~\ref{tab:extendedDatasetExampleCGPT} shows an example conversation generated by ChatGPT while extending the conversation customized by the expert, while Table~\ref{tab:extendedDatasetExampleGemini} presents the same example, but generated by Gemini instead.

\begin{table}[htb]
\centering
\footnotesize
\begin{tabular}{|p{0.1\linewidth}|p{0.8\linewidth}|}
\hline
\textbf{Agent} & \textbf{Utterance} \\ \hline
Speaker & Yesterday I came home and the sink was once again full of dirty dishes. The guy doesn't care. I've already asked him to wash it as he cooks so it doesn't accumulate, but it's no use! I can not take it anymore. \\ \hline

Listener & I'm listening to you and realizing that there's anger, frustration there, does that make sense? \\ \hline

Speaker & Yes, and tiredness as well. \\ \hline

Listener & And perhaps these feelings are present because you need reciprocity and respect. \\ \hline

Speaker & That's it, I really need it. Of consideration as well. \\ \hline

Listener & It sounds like you've been feeling overwhelmed by the situation. \\ \hline

Speaker & Yes, exactly. I feel like I'm the only one taking care of things in the house. \\ \hline

Listener & It must be exhausting to constantly feel like you're carrying all the responsibility. \\ \hline

Speaker & It really is. I just wish he would listen and understand how I'm feeling. \\ \hline

Listener & It sounds like you're seeking understanding and support from him. \\ \hline

Speaker & Yes, that's exactly it. I feel like I'm not being heard. \\ \hline

Listener & It's important to feel heard and validated in your feelings. \\ \hline
\end{tabular}
\caption{Example of conversation generated by ChatGPT, extending the interaction from the expert dataset.}
\label{tab:extendedDatasetExampleCGPT}
\end{table}

\begin{table}[htb]
\centering
\footnotesize
\begin{tabular}{|p{0.1\linewidth}|p{0.8\linewidth}|}
\hline
\textbf{Agent} & \textbf{Utterance} \\ \hline
Speaker & Yesterday I came home and the sink was once again full of dirty dishes. The guy doesn't care. I've already asked him to wash it as he cooks so it doesn't accumulate, but it's no use! I can not take it anymore. \\ \hline

Listener & I'm listening to you and realizing that there's anger, frustration there, does that make sense? \\ \hline

Speaker & Yes, and tiredness as well. \\ \hline

Listener & And perhaps these feelings are present because you need reciprocity and respect. \\ \hline

Speaker & That's it, I really need it. Of consideration as well. \\ \hline

Listener & I can only imagine.  It's disheartening when you express a need and it's not met, especially in a shared living space where cooperation is essential.  It makes you feel unseen and unheard, right? \\ \hline

Speaker & Exactly! It's like my efforts aren't valued, and I'm carrying the weight of the household chores alone. \\ \hline

Listener & So this isn't just about dirty dishes; it's about feeling undervalued and unsupported in the relationship.  That makes perfect sense. It's understandable that you're feeling overwhelmed and resentful. \\ \hline

Speaker & Yes, it's more than just the dishes.  It's the feeling that my concerns are constantly dismissed. It's a pattern that's been happening for a while now. \\ \hline

Listener & It sounds like you’ve been carrying a heavy emotional burden for some time. Have you considered talking to him directly about how this makes you feel, focusing on your feelings rather than blaming? \\ \hline

Speaker & I have tried, but he just says he'll do better and then it happens again. I don't know if it's worth trying again.  I'm starting to feel hopeless. \\ \hline

Listener & I hear that hopelessness.  It’s understandable to feel that way after repeated attempts with little to no change.  Have you considered perhaps exploring other ways to address this, maybe with a therapist or counselor who could help you both communicate more effectively?  Or perhaps even setting clear boundaries? \\ \hline
\end{tabular}
\caption{Example of conversation generated by Gemini, extending the interaction from the expert dataset.}
\label{tab:extendedDatasetExampleGemini}
\end{table}

In its turn, the control datasets (for both ChatGPT and Gemini) were generated to be conversations which may or may not have empathic components (i.e. there was no specific configuration regarding empathetic listening). Fully generated through the LMM, these conversations follow the problem and outcome model similar to empathic listening. It is composed of 30 texts, and, for normalization, only the 12 first interactions of each text were taken into account for the analysis. Table~\ref{tab:controlDatasetExampleCGPT} shows an example conversation fully generated by ChatGPT with no previous configuration concerning empathetic listening, while Table~\ref{tab:controlDatasetExampleGemini} presents the same example, but generated by Gemini instead.

\begin{table}[htb]
\centering
\footnotesize
\begin{tabular}{|p{0.1\linewidth}|p{0.8\linewidth}|}
\hline
\textbf{Agent} & \textbf{Utterance} \\ \hline
Speaker & I just finished a really good book, it was so captivating. \\ \hline

Listener & That's awesome! What was the book about? \\ \hline

Speaker & It was a mystery thriller with a plot twist I never saw coming. \\ \hline

Listener & I love when books surprise me like that. It's such a good feeling. \\ \hline

Speaker & Definitely, it keeps you on your toes. Have you read any good books lately? \\ \hline

Listener & Not recently, I've been so busy with work that I haven't had much time for reading. \\ \hline

Speaker & I totally get that, sometimes life just gets in the way of our hobbies. \\ \hline

Listener & Yeah, but it's important to make time for things we enjoy too. \\ \hline

Speaker & You're right, I need to prioritize some me-time soon. \\ \hline

Listener & You deserve it! What's something you like to do for yourself when you have free time? \\ \hline

Speaker & I love going for long walks in nature, it really helps me clear my mind. \\ \hline

Listener & That sounds so relaxing. I should try that sometime. \\ \hline
\end{tabular}
\caption{Example of conversation fully generated by ChatGPT, with no previous configuration concerning empathetic listening.}
\label{tab:controlDatasetExampleCGPT}
\end{table}

\begin{table}[htb]
\centering
\footnotesize
\begin{tabular}{|p{0.1\linewidth}|p{0.8\linewidth}|}
\hline
\textbf{Agent} & \textbf{Utterance} \\ \hline
Speaker & My dog, Max, finally stopped chewing on my shoes! It's been a battle for months. \\ \hline

Listener & Oh my gosh, that's amazing!  What a relief.  Mine still manages to find new and creative ways to destroy things, though. \\ \hline

Speaker & I know, right? It felt like a victory.  I think it's finally understood "no" means something.  What's your dog getting into these days? \\ \hline

Listener & Mostly the trash.  He's a champion garbage diver.  I've tried everything, but he's relentless.  Speaking of victories, I finally got my car fixed after that fender bender. \\ \hline

Speaker & Oh, good!  What happened?  And what was wrong with it? \\ \hline

Listener & Some lady backed into me at the grocery store.  It was mostly cosmetic damage, but the bumper needed replacing.  It took forever to get the insurance sorted out. \\ \hline

Speaker & That's a pain.  Insurance companies are the worst. At least it's fixed now, though! Anything fun planned for the weekend? \\ \hline

Listener & Not much, just relaxing mostly.  Maybe a movie marathon. How about you? \\ \hline

Speaker & Similar plan, some Netflix and chill.  Maybe I'll finally get around to tackling that book I've been meaning to read. \\ \hline

Listener & Sounds nice and relaxing. I'm looking forward to a quiet weekend after this crazy week.  Getting some well-deserved rest. \\ \hline

Speaker & Totally agree.  We deserve it! I'm already feeling less stressed just thinking about it.  Well, I should get going, got a pile of laundry calling my name. \\ \hline

Listener & Haha, same here!  Talk to you soon. \\ \hline
\end{tabular}
\caption{Example of conversation fully generated by Gemini, with no previous configuration concerning empathetic listening.}
\label{tab:controlDatasetExampleGemini}
\end{table}

In this experiment, our hypothesis is that, when analyzing each interaction between the listener and the speaker, it will be possible to observe a tendency of energy going from negative (or lower values) to positive (or higher values), containing these distinct characteristics:

\begin{itemize}
    \item The conversation begins with the problem, usually negative;
    \item The empathic agent reiterates the problem to make sure he or she understands, equalizing the energy level of the conversation;
    \item The conversation progresses toward a positive conclusion, converging the energy of both agents.
\end{itemize}

In Section~\ref{sec:resultados}, we present the analysis of this experiment.

\subsection{Expert Questionnaire}
\label{sec:expertQuest}

In addition to the lexical analysis, we also asked the expert to evaluate the interactions generated for both datasets (i.e., extended and control) by both chatbots (i.e., ChatGPT and Gemini), resulting in a total of four datasets to be evaluated.
To do so, the expert completed a form, for each of the four datasets, comprising the questions presented in Table~\ref{tab:expertForm}, which is an adaptation of the Bartneck questionnaire~\cite {bartneck2008measuring} with one additional question regarding empathetic listening ability. Moreover, we added a free-text field, allowing the expert to discuss any additional topics not covered by the questions. It is important to clarify that the expert was well aware of the conditions of the experiment, which means he/she knew that the conversations of those datasets were generated by an LLM chatbot. 

\begin{table}[htb]
\centering
\footnotesize
\begin{tabular}{|p{0.8\linewidth}|p{0.1\linewidth}|}
\hline
\textbf{Question} & \textbf{Scale} \\ \hline
Looking at how the conversation evolved, rate your impression of the chatbot on these scales, from Unfriendly to Friendly & 1-5 \\ \hline

Looking at how the conversation evolved, rate your impression of the chatbot on these scales, from Awful to Nice & 1-5 \\ \hline

Looking at how the conversation evolved, rate your impression of the chatbot on these scales, from Foolish to Sensible & 1-5 \\ \hline

How do you evaluate the empathetic listening ability of the chatbot, from Very unsatisfied to Very satisfied? & 1-5 \\ \hline
\end{tabular}
\caption{Questions of the expert form.}
\label{tab:expertForm}
\end{table}

\section{Results}
\label{sec:resultados}

This section presents the results achieved by this work. Section~\ref{sec:expertDataset} includes the analysis of the dataset with the conversations manually tailored by the expert. Section~\ref{sec:extendedDataset} presents the analysis of the extended datasets, which were generated by extending the conversations present in the expert dataset by both ChatGPT and Gemini. Section~\ref{sec:controlDataset} shows the analysis of the control datasets, which were fully generated by both ChatGPT and Gemini to be conversations without LLM training, which could or not present empathetic listening aspects. Finally, Section~\ref{sec:expertEvaluation} includes the evaluation performed by the expert in all datasets.

\subsection{Expert Dataset}
\label{sec:expertDataset}

\begin{figure}[htb]
    \centering
    \subfigure[Energy tendency of the speaker.]{\includegraphics[width=0.49\linewidth]{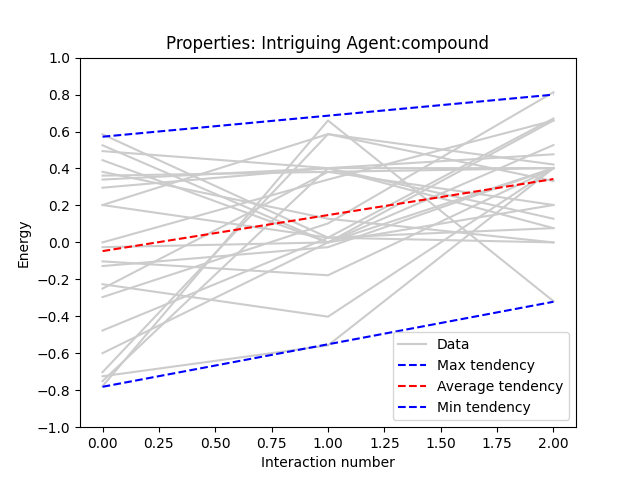}}\hfill
    \subfigure[Energy tendency of the listener.]{\includegraphics[width=0.49\linewidth]{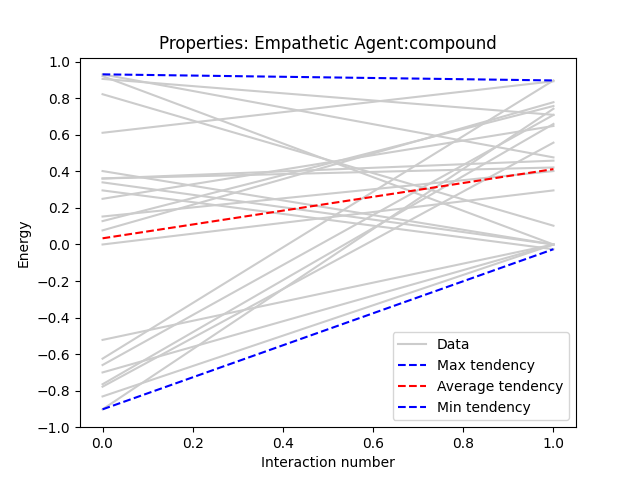}}
    \caption{Energy tendency of the compounded interactions. Y axis shows the energy value (as computed by VADER tool), and X axis shows the flow of the interactions (from the beginning to the end). Blue lines show the maximum and minimum energy tendencies, while the red line represents the average energy tendency. It is possible to notice a growing pattern of the energy tendency, from lower to higher values, as the conversation flowed, which is particularly evident in the red line (Average tendency).}
    \label{fig:expertDataset}
\end{figure}

As commented in Section~\ref{sec:datasets}, a dataset was generated with the conversations that were manually created by the expert, where the listener acted as an empathetic listener. Figure~\ref{fig:expertDataset} shows the energy tendency of the compound of all 24 interactions, for both the speaker (Intriguing Agent, left) and listener (Empathetic Agent, right). The Y axis shows the energy value of the conversation (as computed by VADER tool), and X axis shows the flow of the interactions (from the beginning to the end). Blue lines show the maximum and minimum energy tendencies, while the red line represents the average energy tendency. It is also important to notice that all figures start in 0 for X axis, which means that 0 is the first interaction. It is possible to notice a growing pattern of the energy tendency, from lower to higher values, as the conversation flowed, for both agents, which is particularly evident in the red line of the figures (Average tendency). 

It is interesting to note that this dataset, since it was created by the expert, can also serve as a model, which means that similar patterns should arise from other datasets containing empathetic listening. Thus, the growing pattern (i.e. going from lower to higher energy values) of the Average Tendency (i.e. red line of the figures) should arise as well.

\subsection{Extended Dataset}
\label{sec:extendedDataset}

\begin{figure}[htb]
    \centering
    \subfigure[Energy tendency of the speaker.]{\includegraphics[width=0.49\linewidth]{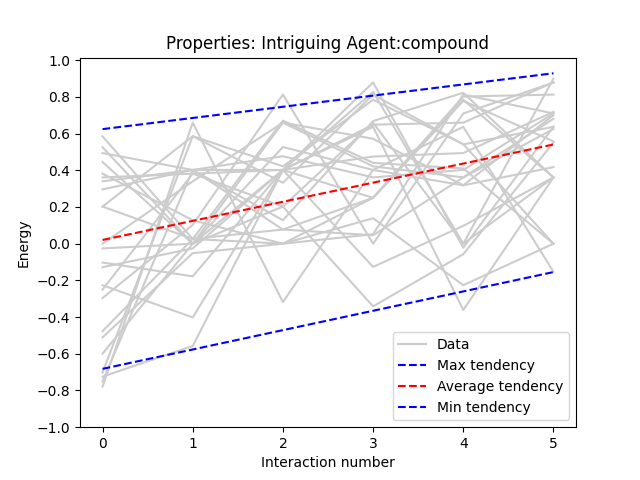}}\hfill
    \subfigure[Energy tendency of the listener.]{\includegraphics[width=0.49\linewidth]{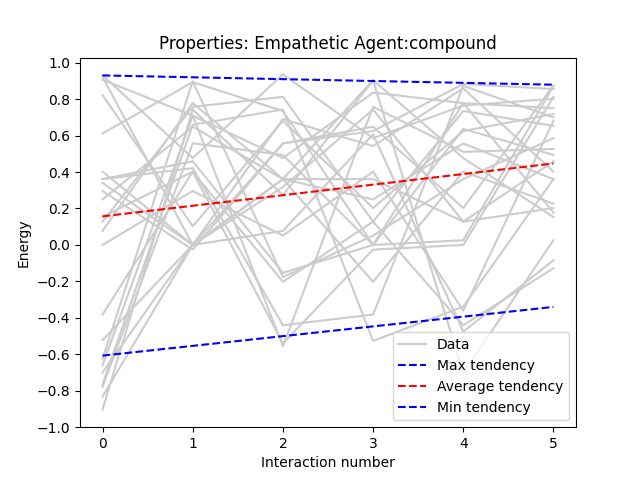}}
    \caption{Energy tendency of the compounded interactions. The Y axis shows the energy value of the conversation (as computed by VADER tool), and X axis shows the flow of the interactions (from the beginning to the end). Blue lines show the maximum and minimum energy tendencies, while the red line represents the average energy tendency. It is possible to notice the continuation of the tendencies generated by the expert dataset, which is particularly evident in the red line of the figures (Average tendency).}
    \label{fig:extendedDatasetLeo}
\end{figure}

As commented in Section~\ref{sec:datasets}, two datasets were generated by extending the conversations of the expert dataset, using both ChatGPT and Gemini. Figure~\ref{fig:extendedDatasetLeo} shows the energy tendency of the compound of all interactions generated by ChatGPT, for both the speaker (Intriguing Agent, left) and listener (Empathetic Agent, right). The Y axis shows the energy value of the conversation (as computed by VADER tool), and X axis shows the flow of the interactions (from the beginning to the end). Blue lines show the maximum and minimum energy tendencies, while the red line represents the average energy tendency. Since the first two points were preserved, the figure shows only 5 interactions for each, speaker and listener. It is possible to notice the continuation of the tendencies generated by the expert dataset, (i.e. a maintenance of the growing pattern), which is particularly evident in the red line of the figures (Average tendency).

\begin{figure}[htb]
    \centering
    \subfigure[Energy tendency of the speaker.]{\includegraphics[width=0.49\linewidth]{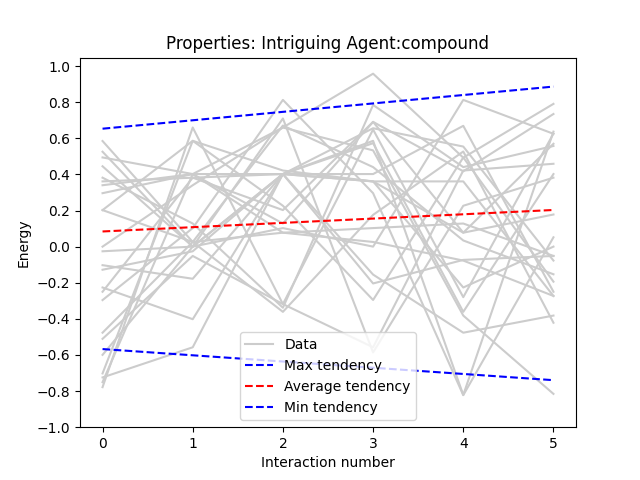}}\hfill
    \subfigure[Energy tendency of the listener.]{\includegraphics[width=0.49\linewidth]{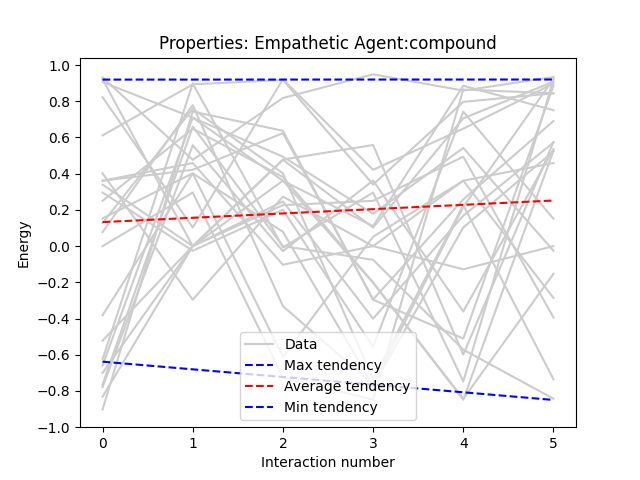}}
    \caption{Energy tendency of the compounded interactions. Y axis shows the energy value (as computed by VADER tool), and X axis shows the flow of the interactions (from the beginning to the end). It is possible to notice the continuation of the tendencies generated by the expert dataset, which is particularly evident in the red line of the figures (Average tendency). It is also possible to notice a reduction on the minimum tendency of both the speaker(a) and the listener(b).}
    \label{fig:extendedDatasetGemini}
\end{figure}

In its turn, Figure~\ref{fig:extendedDatasetGemini} shows the energy tendency of the compound of all interactions generated by Gemini, for both the speaker (Intriguing Agent, left) and listener (Empathetic Agent, right). As it was observed in the extended dataset generated with ChatGPT, it is possible to notice the continuation of the tendencies generated by the expert dataset, which is particularly evident in the red line of the figures (Average tendency). 
However, it is also possible to notice a reduction in the minimum tendency of the speaker~\ref{fig:extendedDatasetGemini}(a) and the listener~\ref{fig:extendedDatasetGemini}(b), which might indicate a difficulty of Gemini in dealing with an empathetic listener role. This insight would be aligned with the expert evaluation, which is presented later in Section~\ref{sec:expertEvaluation}.

\subsection{Control Dataset}
\label{sec:controlDataset}

\begin{figure}[htb]
    \centering
    \subfigure[Energy tendency of the speaker.]{\includegraphics[width=0.49\linewidth]{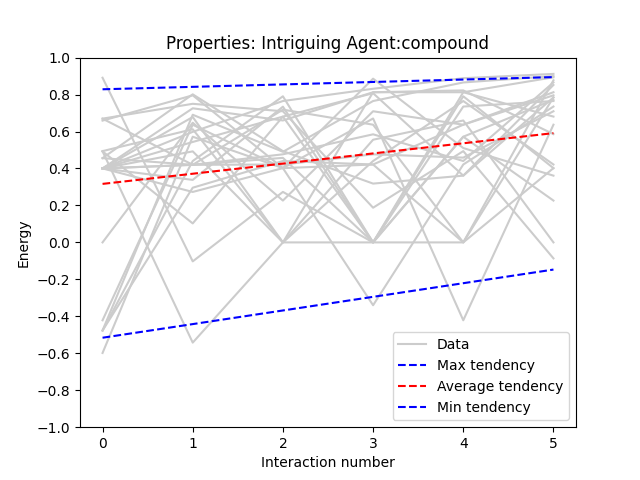}}\hfill
    \subfigure[Energy tendency of the listener.]{\includegraphics[width=0.49\linewidth]{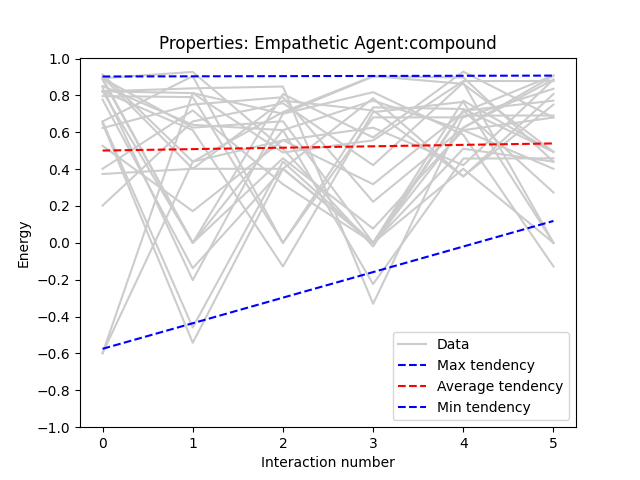}}
    \caption{Energy tendency of the compounded interactions. The Y axis shows the energy value of the conversation (as computed by VADER tool), and X axis shows the flow of the interactions (from the beginning to the end). Blue lines show the maximum and minimum energy tendencies, while the red line represents the average energy tendency. It is possible to notice a difference between both agents, which is particularly evident in the red line of the figures (Average tendency), since the listener (Empathetic Agent(b)) maintained a steady energy during the interactions.}
    \label{fig:controlDatasetLeo}
\end{figure}

As commented in Section~\ref{sec:datasets}, two datasets were generated to be conversations with no particular configuration regarding empathetic listening, using both ChatGPT and Gemini. Figure~\ref{fig:controlDatasetLeo} shows the energy tendency of the compound of all interactions generated by ChatGPT, for both the speaker (Intriguing Agent, left) and listener (Empathetic Agent, right). The Y axis shows the energy value of the conversation (as computed by VADER tool), and X axis shows the flow of the interactions (from the beginning to the end). Blue lines show the maximum and minimum energy tendencies, while the red line represents the average energy tendency. It is possible to notice a difference between both agents, which is particularly evident in the red line of the figures (Average tendency). While the speaker (Intriguing Agent) had a noticeable increase of its energy~\ref{fig:controlDatasetLeo}(a), the listener (Empathetic Agent) had a steady energy~\ref{fig:controlDatasetLeo}(b), showing that its energy remained almost the same for the whole interaction.

\begin{figure}[htb]
    \centering
    \subfigure[Energy tendency of the speaker.]{\includegraphics[width=0.49\linewidth]{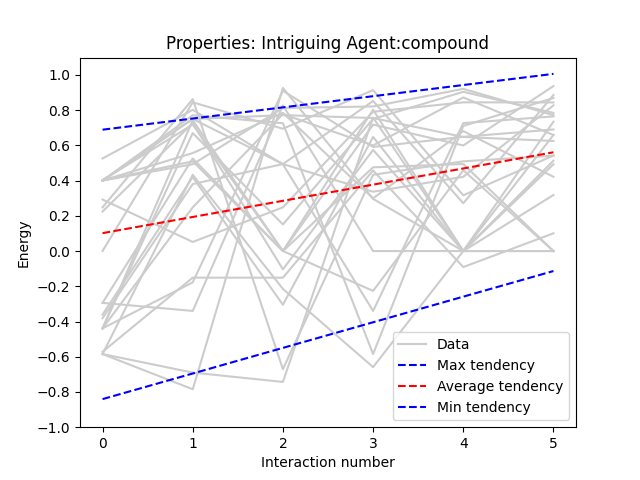}}\hfill
    \subfigure[Energy tendency of the listener.]{\includegraphics[width=0.49\linewidth]{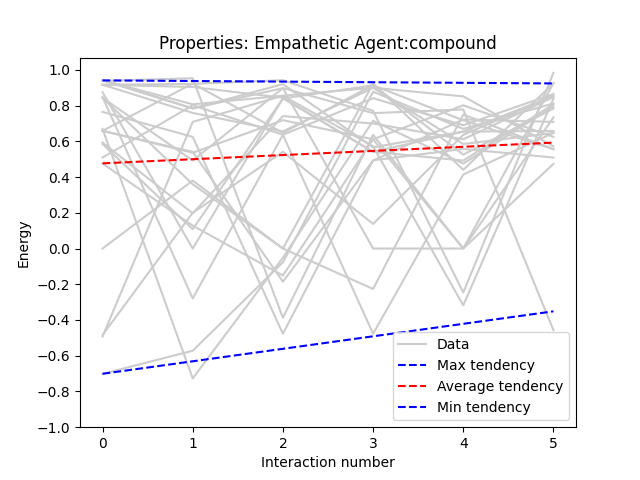}}
    \caption{Energy tendency of the compounded interactions. The Y axis shows the energy value of the conversation (as computed by VADER tool), and X axis shows the flow of the interactions (from the beginning to the end). Blue lines show the maximum and minimum energy tendencies, while the red line represents the average energy tendency. It is possible to notice a difference between both agents, where the listener's energy remained almost stable.}
    \label{fig:controlDatasetGemini}
\end{figure}

Figure~\ref{fig:controlDatasetGemini} shows the energy tendency of the compound of all interactions generated by Gemini, for both the speaker (Intriguing Agent, left) and listener (Empathetic Agent, right). The Y axis shows the energy value of the conversation (as computed by VADER tool), and X axis shows the flow of the interactions (from the beginning to the end). Blue lines show the maximum and minimum energy tendencies, while the red line represents the average energy tendency. As it was observed in the control dataset generated with ChatGPT, it is possible to notice a difference between both agents, which is particularly evident in the red line of the figures (Average tendency). While the speaker had a noticeable increase of its energy~\ref{fig:controlDatasetGemini}(a), the listener had a little increase of its energy~\ref{fig:controlDatasetGemini}(b), showing that its energy remained almost the same for the whole interaction.

After collecting data from the datasets, the results seem to validate our hypothesis: when applying the empathetic listening fine-tuning, a tendency of energy going from lower to higher values could be observed, while conversations seem to have evolved to a positive conclusion, converging the energy of both agents. In its turn, when the fine-tuning was not applied (control datasets), the energy of the listener (Empathetic Agent) remained stable.

\subsection{Expert Evaluation}
\label{sec:expertEvaluation}

As described in Section~\ref{sec:expertQuest}, we asked the expert to evaluate the interactions generated for the datasets. In order to do so, the expert answered a few questions about each dataset, as presented in Table~\ref{tab:expertForm}.
For the first three questions (Friendly, Nice, and Sensible), the expert gave the maximum evaluation (i.e. 5, Very Satisfied) to all four datasets (i.e. Extended and Control, for both ChatGPT and Gemini). However, for the last question (Empathetic listening ability), the expert rated the chatbot with the worst evaluation (i.e. 1, Very Unsatisfied), to all datasets as well. 

By the end of the questionnaire, we included a free text box, so the expert would be able to freely talk about his/her impressions of the evaluated interaction, and even justify the evaluations given. For the Extended dataset generated by ChatGPT, the expert commented that "Although we can classify the chatbot as 'friendly', 'nice' and 'sensible', his performance from the point of view of empathic listening was disastrous. The length of the conversation proposed by the chatbot is very judgmental and does little to explore the feelings and needs of the interlocutor.". For the Extended dataset generated by Gemini, the expert commented that "The extent of the conversation generated by the chatbot is full of judgments and is not restricted to investigating feelings and needs, which would be the function of empathic listening.".

In its turn, for the Control dataset generated by ChatGPT, the expert commented that "In the dialogues created by AI in this case there is no investigation whatsoever by any of the interlocutor's feelings and needs.". In its turn, for the Control dataset generate by Gemini, the expert commented that "There is no interest whatsoever on the part of the interlocutors in the feelings and needs of the other, which would be the main characteristic of empathic listening.".

The results achieved with the energy tendency reading and with the expert evaluation, for both datasets (Extended and Control) and for both chatbots (ChatGPT and Gemini), suggest that, although useful, a lexical analysis should not be used as the only evaluation metric of an empathetic listener chatbot. Even when the energy of the conversation rises over time, the interaction itself might not be beneficial to the user, especially if the chatbot, who should be acting as an empathetic listener, does not fill this role entirely. In fact, the feedback of the expert suggests that the LLM-powered chatbot was not able to act as an empathetic listener, given its judgmental comments and lack of feeling exploring.

On the other hand, by the comments provided, it is possible to notice that the fine-tuned chatbot tried to fill the empathetic listener role, but failed, as the expert commented that it was judgmental and did little to explore the user's feeling and needs. In its turn, the chatbot with no fine-tuning didn't even try to fill the role, as the expert commented that there was no investigation about the feelings and needs of the user.

\section{Final Considerations}
\label{sec:conclusao}




This study evaluated whether fine-tuning large language models can improve their ability to engage in empathetic listening dialogue. Our experiments involved ChatGPT and Gemini, both adapted through prompt engineering and assessed using energy trajectory analysis and expert feedback.

The obtained results reinforce the relevance of using both human-centered and automated approaches to analyze emotional progressions in dialogue. Beyond that, the study highlights two key takeaways. First, it was encouraging to observe that LLMs can be prompted and guided toward producing more emotionally supportive responses, showing potential for fine-tuning in the direction of empathy. However, the second—and perhaps more critical—finding is that truly empathetic listening goes beyond generating positive or soothing messages. It involves engaging with the speaker’s emotional state, helping them reflect and better understand their own problems—something that current LLMs still struggle to achieve. Finally, while sentiment lexicons such as VADER can be helpful tools for analyzing emotional trends, they alone are not sufficient to evaluate the presence or quality of empathetic listening. Empathy in dialogue is a complex, multi-layered process that requires context, intention, and human sensitivity—dimensions that remain challenging for automated systems.


From a human-centered AI perspective, this research reinforces the value of tailoring dialogue systems for affective sensitivity. Empathetic listening is critical in areas such as mental health support, digital companionship, and training simulations. Our methodology highlights the importance of expert-curated datasets and emotional metrics in evaluating conversational empathy.

For future work, we propose extending this investigation to additional LLMs, such as LLaMA or DeepSeek, to assess generalization. Moreover, involving interdisciplinary experts—including psychologists—and conducting user-based studies would enrich both the technical and social understanding of empathetic virtual agents.

\begin{acks}
This study was partly financed by the Coordenação de Aperfeiçoamento de Pessoal de Nivel Superior – Brazil (CAPES) – Finance Code 001; by the Conselho Nacional de Desenvolvimento Científico e Tecnológico - Brazil (CNPq) - Process Numbers 309228/2021-2; 406463/2022-0; 153641/2024-0.
\end{acks}

\bibliographystyle{ACM-Reference-Format}
\bibliography{sample-base}


\begin{thebibliography}{14}


\ifx \showCODEN    \undefined \def \showCODEN     #1{\unskip}     \fi
\ifx \showISBNx    \undefined \def \showISBNx     #1{\unskip}     \fi
\ifx \showISBNxiii \undefined \def \showISBNxiii  #1{\unskip}     \fi
\ifx \showISSN     \undefined \def \showISSN      #1{\unskip}     \fi
\ifx \showLCCN     \undefined \def \showLCCN      #1{\unskip}     \fi
\ifx \shownote     \undefined \def \shownote      #1{#1}          \fi
\ifx \showarticletitle \undefined \def \showarticletitle #1{#1}   \fi
\ifx \showURL      \undefined \def \showURL       {\relax}        \fi
\providecommand\bibfield[2]{#2}
\providecommand\bibinfo[2]{#2}
\providecommand\natexlab[1]{#1}
\providecommand\showeprint[2][]{arXiv:#2}

\bibitem[Abd-Alrazaq et~al\mbox{.}(2023)]%
        {abd2023large}
\bibfield{author}{\bibinfo{person}{Alaa Abd-Alrazaq}, \bibinfo{person}{Rawan AlSaad}, \bibinfo{person}{Dari Alhuwail}, \bibinfo{person}{Arfan Ahmed}, \bibinfo{person}{Padraig~Mark Healy}, \bibinfo{person}{Syed Latifi}, \bibinfo{person}{Sarah Aziz}, \bibinfo{person}{Rafat Damseh}, \bibinfo{person}{Sadam~Alabed Alrazak}, \bibinfo{person}{Javaid Sheikh}, {et~al\mbox{.}}} \bibinfo{year}{2023}\natexlab{}.
\newblock \showarticletitle{Large language models in medical education: opportunities, challenges, and future directions}.
\newblock \bibinfo{journal}{\emph{JMIR Medical Education}} \bibinfo{volume}{9}, \bibinfo{number}{1} (\bibinfo{year}{2023}), \bibinfo{pages}{e48291}.
\newblock


\bibitem[Arjmand et~al\mbox{.}(2024)]%
        {arjmand2024empathic}
\bibfield{author}{\bibinfo{person}{Mehdi Arjmand}, \bibinfo{person}{Farnaz Nouraei}, \bibinfo{person}{Ian Steenstra}, {and} \bibinfo{person}{Timothy Bickmore}.} \bibinfo{year}{2024}\natexlab{}.
\newblock \showarticletitle{Empathic Grounding: Explorations using Multimodal Interaction and Large Language Models with Conversational Agents}. In \bibinfo{booktitle}{\emph{Proceedings of the ACM International Conference on Intelligent Virtual Agents}}. \bibinfo{pages}{1--10}.
\newblock


\bibitem[Bartneck et~al\mbox{.}(2008)]%
        {bartneck2008measuring}
\bibfield{author}{\bibinfo{person}{Christoph Bartneck}, \bibinfo{person}{Elizabeth Croft}, {and} \bibinfo{person}{Dana Kulic}.} \bibinfo{year}{2008}\natexlab{}.
\newblock \showarticletitle{Measuring the anthropomorphism, animacy, likeability, perceived intelligence and perceived safety of robots}.
\newblock  (\bibinfo{year}{2008}).
\newblock


\bibitem[Bill and Eriksson(2023)]%
        {bill2023fine}
\bibfield{author}{\bibinfo{person}{Desir{\'e}e Bill} {and} \bibinfo{person}{Theodor Eriksson}.} \bibinfo{year}{2023}\natexlab{}.
\newblock \bibinfo{title}{Fine-tuning a llm using reinforcement learning from human feedback for a therapy chatbot application}.
\newblock


\bibitem[Brown et~al\mbox{.}(2020)]%
        {brown2020language}
\bibfield{author}{\bibinfo{person}{Tom Brown}, \bibinfo{person}{Benjamin Mann}, \bibinfo{person}{Nick Ryder}, \bibinfo{person}{Melanie Subbiah}, \bibinfo{person}{Jared~D Kaplan}, \bibinfo{person}{Prafulla Dhariwal}, \bibinfo{person}{Arvind Neelakantan}, \bibinfo{person}{Pranav Shyam}, \bibinfo{person}{Girish Sastry}, \bibinfo{person}{Amanda Askell}, {et~al\mbox{.}}} \bibinfo{year}{2020}\natexlab{}.
\newblock \showarticletitle{Language models are few-shot learners}.
\newblock \bibinfo{journal}{\emph{Advances in neural information processing systems}}  \bibinfo{volume}{33} (\bibinfo{year}{2020}), \bibinfo{pages}{1877--1901}.
\newblock


\bibitem[Chen et~al\mbox{.}({[n.\,d.]})]%
        {chen2023soulchat}
\bibfield{author}{\bibinfo{person}{Yirong Chen}, \bibinfo{person}{Xiaofen Xing}, \bibinfo{person}{Jingkai Lin}, \bibinfo{person}{Zhenyu Wang}, \bibinfo{person}{Qi Liu}, \bibinfo{person}{Xiangmin Xu}, {et~al\mbox{.}}} \bibinfo{year}{[n.\,d.]}\natexlab{}.
\newblock \showarticletitle{SoulChat: Improving LLMs' Empathy, Listening, and Comfort Abilities through Fine-tuning with Multi-turn Empathy Conversations}. In \bibinfo{booktitle}{\emph{The 2023 Conference on Empirical Methods in Natural Language Processing}}.
\newblock


\bibitem[Elyoseph et~al\mbox{.}(2023)]%
        {elyoseph2023chatgpt}
\bibfield{author}{\bibinfo{person}{Zohar Elyoseph}, \bibinfo{person}{Dorit Hadar-Shoval}, \bibinfo{person}{Kfir Asraf}, {and} \bibinfo{person}{Maya Lvovsky}.} \bibinfo{year}{2023}\natexlab{}.
\newblock \showarticletitle{ChatGPT outperforms humans in emotional awareness evaluations}.
\newblock \bibinfo{journal}{\emph{Frontiers in Psychology}}  \bibinfo{volume}{14} (\bibinfo{year}{2023}), \bibinfo{pages}{1199058}.
\newblock


\bibitem[Hasan et~al\mbox{.}(2023)]%
        {hasan2023sapien}
\bibfield{author}{\bibinfo{person}{Masum Hasan}, \bibinfo{person}{Cengiz Ozel}, \bibinfo{person}{Sammy Potter}, {and} \bibinfo{person}{Ehsan Hoque}.} \bibinfo{year}{2023}\natexlab{}.
\newblock \showarticletitle{SAPIEN: affective virtual agents powered by large language models}. In \bibinfo{booktitle}{\emph{2023 11th International Conference on Affective Computing and Intelligent Interaction Workshops and Demos (ACIIW)}}. IEEE, \bibinfo{pages}{1--3}.
\newblock


\bibitem[Hutto and Gilbert(2014)]%
        {hutto2014vader}
\bibfield{author}{\bibinfo{person}{Clayton Hutto} {and} \bibinfo{person}{Eric Gilbert}.} \bibinfo{year}{2014}\natexlab{}.
\newblock \showarticletitle{Vader: A parsimonious rule-based model for sentiment analysis of social media text}. In \bibinfo{booktitle}{\emph{Proceedings of the international AAAI conference on web and social media}}, Vol.~\bibinfo{volume}{8}. \bibinfo{pages}{216--225}.
\newblock


\bibitem[Manickam et~al\mbox{.}(2022)]%
        {manickam2022artificial}
\bibfield{author}{\bibinfo{person}{Pandiaraj Manickam}, \bibinfo{person}{Siva~Ananth Mariappan}, \bibinfo{person}{Sindhu~Monica Murugesan}, \bibinfo{person}{Shekhar Hansda}, \bibinfo{person}{Ajeet Kaushik}, \bibinfo{person}{Ravikumar Shinde}, {and} \bibinfo{person}{SP Thipperudraswamy}.} \bibinfo{year}{2022}\natexlab{}.
\newblock \showarticletitle{Artificial intelligence (AI) and internet of medical things (IoMT) assisted biomedical systems for intelligent healthcare}.
\newblock \bibinfo{journal}{\emph{Biosensors}} \bibinfo{volume}{12}, \bibinfo{number}{8} (\bibinfo{year}{2022}), \bibinfo{pages}{562}.
\newblock


\bibitem[Prendinger et~al\mbox{.}(2005)]%
        {prendinger2005using}
\bibfield{author}{\bibinfo{person}{Helmut Prendinger}, \bibinfo{person}{Junichiro Mori}, {and} \bibinfo{person}{Mitsuru Ishizuka}.} \bibinfo{year}{2005}\natexlab{}.
\newblock \showarticletitle{Using human physiology to evaluate subtle expressivity of a virtual quizmaster in a mathematical game}.
\newblock \bibinfo{journal}{\emph{International journal of human-computer studies}} \bibinfo{volume}{62}, \bibinfo{number}{2} (\bibinfo{year}{2005}), \bibinfo{pages}{231--245}.
\newblock


\bibitem[Talyshinskii et~al\mbox{.}(2024)]%
        {talyshinskii2024potential}
\bibfield{author}{\bibinfo{person}{Ali Talyshinskii}, \bibinfo{person}{Nithesh Naik}, \bibinfo{person}{BM~Zeeshan Hameed}, \bibinfo{person}{Patrick Julieb{\o}-Jones}, {and} \bibinfo{person}{Bhaskar~Kumar Somani}.} \bibinfo{year}{2024}\natexlab{}.
\newblock \showarticletitle{Potential of AI-driven chatbots in urology: revolutionizing patient care through artificial intelligence}.
\newblock \bibinfo{journal}{\emph{Current Urology Reports}} \bibinfo{volume}{25}, \bibinfo{number}{1} (\bibinfo{year}{2024}), \bibinfo{pages}{9--18}.
\newblock


\bibitem[Team et~al\mbox{.}(2023)]%
        {team2023gemini}
\bibfield{author}{\bibinfo{person}{Gemini Team}, \bibinfo{person}{Rohan Anil}, \bibinfo{person}{Sebastian Borgeaud}, \bibinfo{person}{Jean-Baptiste Alayrac}, \bibinfo{person}{Jiahui Yu}, \bibinfo{person}{Radu Soricut}, \bibinfo{person}{Johan Schalkwyk}, \bibinfo{person}{Andrew~M Dai}, \bibinfo{person}{Anja Hauth}, \bibinfo{person}{Katie Millican}, {et~al\mbox{.}}} \bibinfo{year}{2023}\natexlab{}.
\newblock \showarticletitle{Gemini: a family of highly capable multimodal models}.
\newblock \bibinfo{journal}{\emph{arXiv preprint arXiv:2312.11805}} (\bibinfo{year}{2023}).
\newblock


\bibitem[Weizenbaum(1966)]%
        {weizenbaum1966eliza}
\bibfield{author}{\bibinfo{person}{Joseph Weizenbaum}.} \bibinfo{year}{1966}\natexlab{}.
\newblock \showarticletitle{ELIZA—a computer program for the study of natural language communication between man and machine}.
\newblock \bibinfo{journal}{\emph{Commun. ACM}} \bibinfo{volume}{9}, \bibinfo{number}{1} (\bibinfo{year}{1966}), \bibinfo{pages}{36--45}.
\newblock


\end{thebibliography}










\end{document}